\documentclass[%
 aps,
 prl,
 amsmath,amssymb,
 superscriptaddress,
 reprint,%
]{revtex4-2}

\usepackage{comment}
\usepackage{graphicx}
\usepackage{dcolumn}
\usepackage{bm}

\usepackage{xcolor}

\begin{document}

\preprint{AIP/123-QED}

\title[Arbitrarily Structured Laser Pulses]{Arbitrarily Structured Laser Pulses}

\author{Jacob R. Pierce}
    \email{jacobpierce@physics.ucla.edu}
    \affiliation{Department of Physics and Astronomy, University of California, Los Angeles, California 90095, USA}
 
 \author{John P. Palastro}
    \email{jpal@lle.rochester.edu}
    \affiliation{ University of Rochester, Laboratory for Laser Energetics, Rochester, New York 14623, USA}


\author{Fei Li}
    \affiliation{Department of Physics and Astronomy, University of California, Los Angeles, California 90095, USA}

\author{Bernardo Malaca}
    \affiliation{GoLP/Instituto de Plasmas e Fusão Nuclear, Instituto Superior Técnico, Universidade de Lisboa, Lisbon 1049-001, Portugal}

\author{Dillon Ramsey} 
    \affiliation{ University of Rochester, Laboratory for Laser Energetics, Rochester, New York 14623, USA}


\author{Jorge Vieira}
    \affiliation{GoLP/Instituto de Plasmas e Fusão Nuclear, Instituto Superior Técnico, Universidade de Lisboa, Lisbon 1049-001, Portugal}

    
\author{Kathleen Weichman} 
    \affiliation{ University of Rochester, Laboratory for Laser Energetics, Rochester, New York 14623, USA}
    
\author{Warren B. Mori}
    \affiliation{Department of Physics and Astronomy, University of California, Los Angeles, California 90095, USA}


\date{\today}

\begin{abstract}
   
    Spatiotemporal control refers to a class of optical techniques for structuring a laser pulse with coupled space-time dependent properties, including moving focal points, dynamic spot sizes, and evolving orbital angular momenta. Here we introduce the concept of arbitrarily structured laser (ASTRL) pulses which generalizes these techniques. The ASTRL formalism employs a superposition of prescribed pulses to create a desired electromagnetic field structure. Several examples illustrate the versatility of ASTRL pulses to address a range of laser-based applications.
    
\end{abstract}

\maketitle

    Light can exhibit structures in space, time, or coupled space-time. The realization that these structures can be shaped using optical techniques has led to a deeper understanding of the fundamental properties of light and advances in multiple laser-based applications. Spatial shaping involves the use of static optical elements, such as phase plates, deformable mirrors, or spatial light modulators, to form transverse structures with orbital angular momentum, Airy or Bessel profiles, or tailored speckle patterns \cite{beijersbergen1994spiralphase,siviloglou2007airy,chattrapiban2003bessel,kato1984phaseplate}. Temporal shaping typically employs Fourier methods, in which frequency components are separated by diffraction gratings and separately modified, to synthesize pulses with nearly arbitrary time-dependent waveforms and polarizations \cite{weiner2000shaping,weiner2011ultrafast,brixner2001femtosecond}. 

    Spatiotemporal pulse shaping invokes some combination of these techniques. Examples go back as far as smoothing by spectral dispersion (1989)  \cite{skupsky1989improved} and pulse front tilt (1996) \cite{hebling1996derivation}. More modern examples include spatiotemporal optical vortices (STOVs) \cite{jhajj2016spatiotemporal,hancock2019free,hancock2021stov}, diffraction-free light sheets \cite{kondakci2016diffraction,Kondakci2017,kondakci2019optical}, and flying focus concepts \cite{Sainte-Marie2017,Froula2018,Simpson:20,Simpson:22}. Flying focus pulses, in particular, feature intensity peaks with programmable trajectories that can travel distances far greater than a Rayleigh range while maintaining a near-constant profile. The trajectory control afforded by these pulses has been proposed as a means to overcome limitations of traditional pulses in a number of applications, including laser wakefield acceleration \cite{Palastro2020,Palastro2021,Caizergues2020}, Raman amplification \cite{Turnbull2018}, photon acceleration \cite{Howard2019}, vacuum electron acceleration \cite{Ramsey2020}, and nonlinear Thomson scattering \cite{ramsey2022nonlinear}, as well as for the exploration of fundamental physics, such as radiation reaction \cite{Formanek2021} and nonlinear Compton scattering \cite{Piazza2021}.
    
    There are two approaches for creating spatiotemporally structured pulses. The first approach requires identifying a specific solution to Maxwell's equations. Once known, this solution can be synthesized using programmable phase modulation techniques. This is the approach taken to create light sheets and STOVs. However, this approach does not provide a method to systematically identify new solutions of Maxwell's equations with features that are desirable for a particular application. In the second approach, one imagines an electromagnetic structure with desirable features and synthesizes this structure using a superposition of solutions to Maxwell's equations with known properties. This is the approach presented here.
    
    In this letter, we introduce a concept and theoretical formalism for generating laser pulses with arbitrary structure. These arbitrarily structured laser (ASTRL) pulses are synthesized using a superposition of traditional laser pulses with controlled and varying properties. The ASTRL formalism generalizes the creation of pulses with evolving focal points (flying foci), spot size, orbital angular momentum, and polarization, and can also describe pulses with exotic topological structure such as STOVs. An example of each of these is presented. The flexibility of the ASTRL concept opens new and previously unimagined possibilities for structured light with the potential to improve a wide range of laser-based applications, while its simplicity facilitates implementation into simulations. 
    
    The formulation of the ASTRL concept begins with the vacuum wave equation for the electric field of a laser pulse:
        \begin{equation}
    	\left[ \nabla^2 -\frac{1}{c^2}\partial_t^2 \right] \mathbf E(\mathbf x, t)=0.
    	\label{eq:waveequation}
    	\end{equation}
    ASTRL pulses are constructed by superposing solutions to Eq. \ref{eq:waveequation} with varying properties. The electric field of an ASTRL pulse can be expressed either as a sum or an integral over these solutions:  
	\begin{equation}
        \mathbf E(\mathbf x, t) = \int d\eta \; \mathbf E_\eta(\mathbf x,t ), 
		\label{eq:astrl-most-general}
	\end{equation}    
    where $\eta$ parameterizes the varying properties of the solutions. For example the amplitude, spot size, focal point, pulse duration, or relative delay of each $\mathbf E_\eta $ can all depend on $\eta$. Equation \ref{eq:astrl-most-general} is the most general representation of an ASTRL pulse in vacuum. 
    
    Analytic solutions for $\mathbf E_\eta$ simplify the process of designing an ASTRL pulse for a specific application. Paraxial solutions, in particular, are expressed in terms of familiar quantities that can be parameterized in terms of $\eta$. To derive such solutions, consider a laser pulse propagating in the $\hat z$-direction with constant polarization $\hat{ \boldsymbol{ \epsilon} }$. The transverse electric field can be expressed as an envelope $A(\mathbf x_\perp, z, t)$ modulated by a carrier
    	\begin{equation}
    	\mathbf E_\perp(\mathbf x_\perp, z, t) = \tfrac{1}{2}A(\mathbf x_\perp, z, t)e^{i(k_0z-\omega_0t)} \hat{\boldsymbol \epsilon } + \mathrm{c.c.},
    	\label{eq:envelope-def}
	\end{equation}
    where $k_0$ is the central wavenumber and $\omega_0 = ck_0$. Substituting Eq. \ref{eq:envelope-def} into Eq. \ref{eq:waveequation} and performing the Galilean change of variables $(\xi,s)=(z-ct,z)$ provides 
	\begin{equation}
        \left[ \nabla_\perp^2 + \partial_s^2 + 2ik_0 \partial_s + 2 \partial_\xi \partial_s \right] A (\mathbf x_\perp,s,\xi)=0.
        \label{eq:envelope-galilean}
    \end{equation}
    Upon applying the slowly-varying envelope approximation
    ($|\partial_{\xi}|, |\partial_s| \ll k_0$), Eq. \ref{eq:envelope-galilean} reduces to
        \begin{equation}
        \left[ \nabla_\perp^2 + 2ik_0 \partial_s \right] A (\mathbf x_\perp,s,\xi) \approx 0.
        \label{eq:envelope-galilean-approx}
    \end{equation}
    Equation \ref{eq:envelope-galilean-approx} is the paraxial wave equation and its operator is independent of $\xi$. This independence admits separable solutions, i.e., solutions with no spatiotemporal coupling, of the form
	\begin{equation}
	    A(\mathbf x_\perp,s,\xi) \approx B(\xi) C(\mathbf x_\perp,s),
        \label{eq:decomp}
	\end{equation}
	where $C(\mathbf x_\perp,s)$ satisfies $ [ \nabla_\perp^2 + 2 i k_0 \partial_s  ] C(\mathbf x_\perp,s ) = 0 $ and $B (\xi)$ is an arbitrary function. For example, $C(\mathbf x_\perp,s )$ may be any Laguerre-, Hermite-, or Ince-Gaussian mode and the longitudinal profile may be $B(\xi) = \exp[ -(\xi / c\tau_0)^2 ] $. 
	
	The transverse electric field of a coherent ASTRL pulse can be approximated by summing or integrating over paraxial solutions:   
	\begin{equation}
        \mathbf E_\perp(\mathbf x_\perp,s,\xi) \approx \tfrac{1}{2} e^{ik_0\xi} \int d\eta \; B_\eta(\xi) C_\eta(\mathbf x_\perp,s) \hat {\boldsymbol \epsilon}(\eta) + \mathrm{c.c.}, 
		\label{eq:astrl-general}
	\end{equation}      
    where the properties of $B_\eta(\xi)$, $C_\eta(\mathbf{x}_\perp,s)$, and $\hat{\boldsymbol \epsilon}(\eta)$ may vary with respect to $\eta$. The integral over $\eta$ may introduce spatiotemporal coupling, such that in general, Eq. \ref{eq:astrl-general} cannot be written as a separable function. Thus, $\mathbf E_\perp$ may possess a high degree of spatiotemporal coupling despite the absence of coupling in Eq. \ref{eq:decomp}.
    
    The $\eta$-dependence of $B_\eta(\xi)$ may encode a delay $\Delta$, duration $\tau_0$, and weight $B_0(\eta)$ for each constituent pulse as \begin{equation}
	    B_\eta(\xi) = B_0(\eta) \exp \left[ -\left( \frac{ \xi - \Delta(\eta)}{c\tau_0(\eta)} \right) ^2 \right] .   
	    \label{eq:B_delay}
	\end{equation}
    The durations $\tau_0(\eta)$ set the fastest timescale over which the properties of the ASTRL pulse can vary and determine the number of pulses required for discrete approximation of Eq. \ref{eq:astrl-general} (see Supplemental). Unless otherwise stated, $\tau_0$ and $B_0$ will be independent of $\eta$ for the examples presented here.
    
    \begin{figure}[h!]
        \centering
        \includegraphics{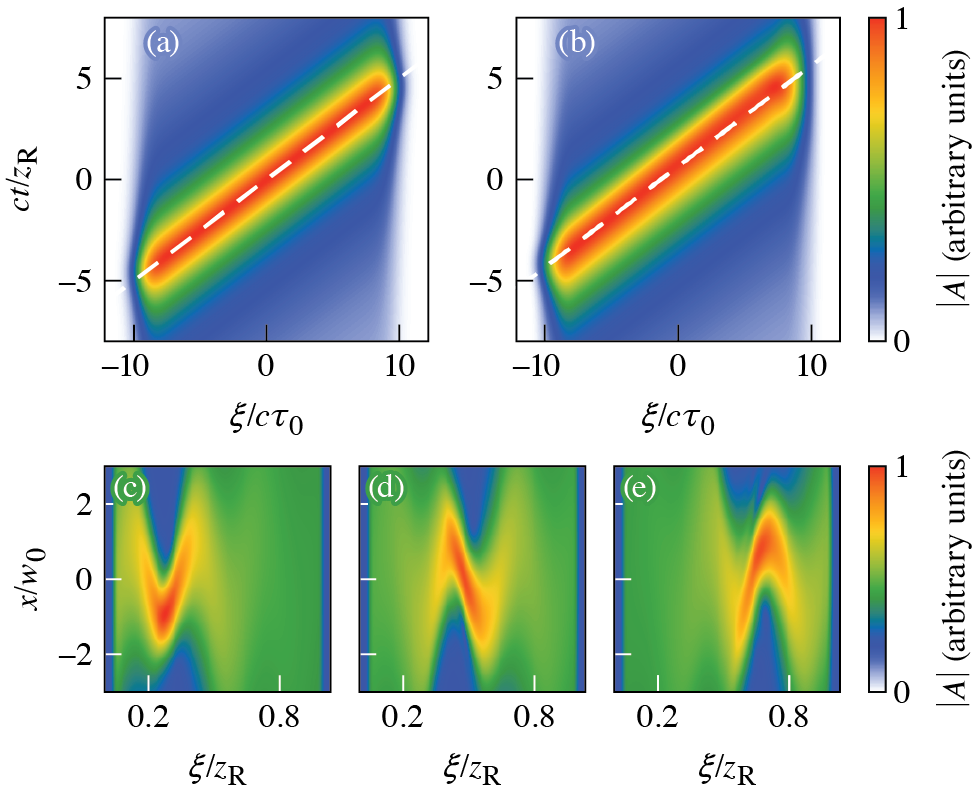}
        \caption{(a): On-axis evolution of the envelope of an ASTRL flying focus pulse found by evaluating Eq. \ref{eq:astrl-general} with $k_0 w_0 = 40 $, $ \omega_0 \tau_0 = 20 $, and $\beta_f = 1.05$. The dashed line indicates the trajectory $z = \beta_f t $. (b) The envelope from a quasi-3D FDTD simulation for the same parameters. The two are in excellent agreement. (c) -- (e): A generalized flying focus constructed using the ASTRL formalism in which the focal point oscillates transversely as it moves along the optical axis. The propagation was simulated in 2D using the FDTD method. The pulse was initialized at $ct/z_R=-10$ using Eq.  \ref{eq:astrl-general} with $x_{\perp\text{osc}} = w_0$, $k_\text{osc} = 1/z_R$, $\beta_f=1.05$, $L=20z_R$,  $k_0w_0 = 40$, and $\omega_0 \tau_0 = 20$. From left to right, the frames show the pulse at $ct/z_R = -4$, $0$, and $4$.} 
        \label{fig:flyfoc}
    \end{figure}
    
	A natural example for the $\eta$-dependence of $C_{\eta}(\mathbf x_\perp,s )$ is an $\eta$-indexed Gaussian-beam:
    \begin{equation}
		C_{\eta}(\mathbf x_\perp,s ) = \left[\frac{z_R(\eta)}{q_\eta(s)}\right]^{d_\perp / 2} \exp\left[ - \frac{ i k_0 |\mathbf x_\perp - \mathbf x_{\perp0}(\eta)|^2}{2 q_\eta(s)}  \right],
		\label{eq:gaussian-beam}
	\end{equation}	
	where $q_\eta(s) \equiv s - s_0(\eta) + iz_R(\eta)$ is the complex beam parameter and $d_{\perp}$ is the number of transverse dimensions. In Eq. \ref{eq:gaussian-beam}, the longitudinal focal position $ s_0(\eta)$, transverse focal position $\mathbf x_{\perp0}(\eta) $, and Rayleigh range $z_R(\eta)$ (and associated spot size $w_0(\eta) = \sqrt{2 z_R(\eta) / k_0}$) may all be expressed as functions of the integration variable $\eta$, which allows for extended focal ranges and evolving transverse structure. More generally, the mode numbers of the transverse profile can depend on $\eta$, e.g., the radial or orbital angular momentum (OAM) quantum numbers of a Laguerre-Gaussian mode. Finally, $\hat{ \boldsymbol \epsilon}(\eta)$ may encode an evolving polarization. 
	
	The utility and generality of the ASTRL concept will be demonstrated by using Eq. \ref{eq:astrl-general} to construct several examples of spatiotemporally structured laser pulses. To begin, consider the flying focus \cite{Sainte-Marie2017,Froula2018}. Each time slice in a flying focus pulse has a different focal point along the propagation axis. The time and location at which each slice comes to focus can be controlled to produce an intensity peak that travels at any velocity, including backwards or faster than the speed of light. Within the ASTRL formalism, a flying focus pulse may be described using a Gaussian transverse profile (Eq. \ref{eq:gaussian-beam}) with $ \eta \in \left[-\frac 1 2, \frac 1 2 \right] $ and $s_0(\eta) = \eta L $. This choice of $s_0(\eta)$ produces a focal range extending from $s=-L/2$ to $s=L/2$, where $L$ can be much larger than $z_R$. At each $s$ in the focal range, the focus will occur at a time determined by the condition $\xi = \Delta(\eta) = \Delta(s/L)$. The delay $\Delta(\eta)$ required for a given focal trajectory can be derived from geometric optics as in Ref \cite{Palastro2020}. For example, motion of the focal point at a constant velocity $\mathrm{v}_f = c\beta_f$ is attained with $\Delta(\eta) = ( 1 - 1 / \beta_f  )L \eta$. 
	
	Figure \ref{fig:flyfoc}(a) displays the on-axis envelope of a superluminal ($\beta_f = 1.05$) flying focus pulse obtained by evaluating Eq. \ref{eq:astrl-general}. 
	The white dashed line shows the spacetime trajectory $ z = \beta_f t $ and demonstrates that the laser envelope travels at the expected velocity over 10 Rayleigh ranges. A finite-difference time-domain (FDTD) simulation of the full set of Maxwell's equations produces nearly identical results [Fig. \ref{fig:flyfoc}(b)]. The simulation used Eq. \ref{eq:astrl-general} to initialize the fields, but did not make the approximations leading to Eq. \ref{eq:envelope-galilean-approx}. A slight discrepancy can be observed near $ct\approx 5z_R$ due to the inexactness of Eq. \ref{eq:decomp}. The simulation was conducted using the code OSIRIS \cite{Fonseca2002} with quasi-3D azimuthal decomposition \cite{Davidson2015} and the solver introduced in Ref \cite{Li2021} to mitigate numerical dispersion. 

    The ASTRL concept can also be used to describe new types of generalized flying focus pulses. For instance, a pulse whose focal point oscillates in the transverse direction as it translates along the propagation axis could serve as a controllable wiggler for generating radiation from relativistic electron bunches or enhance direct laser acceleration of electrons through new parametric resonances.  Such a pulse can be constructed by parameterizing both the transverse and longitudinal focal coordinates: $\mathbf x_{\perp0}(\eta) = x_{\perp\text{osc}} \sin[ k_\text{osc} s_0(\eta)] \hat {\mathbf x}$ and $s_0(\eta) = \eta L $, where, as before, $ \eta \in \left[-\frac 1 2, \frac 1 2 \right] $. Figures \ref{fig:flyfoc}(c)--(e) display the envelope from 2D OSIRIS simulations of this configuration. The intensity peak undergoes the expected oscillatory motion. Here, $C_{\eta}$ is given by Eq. \ref{eq:gaussian-beam} and $\Delta(\eta) = ( 1 - 1 / \beta_f  )L \eta$, i.e., the focal point moves with a constant axial velocity. 
    
\begin{figure}[h!]
         \includegraphics{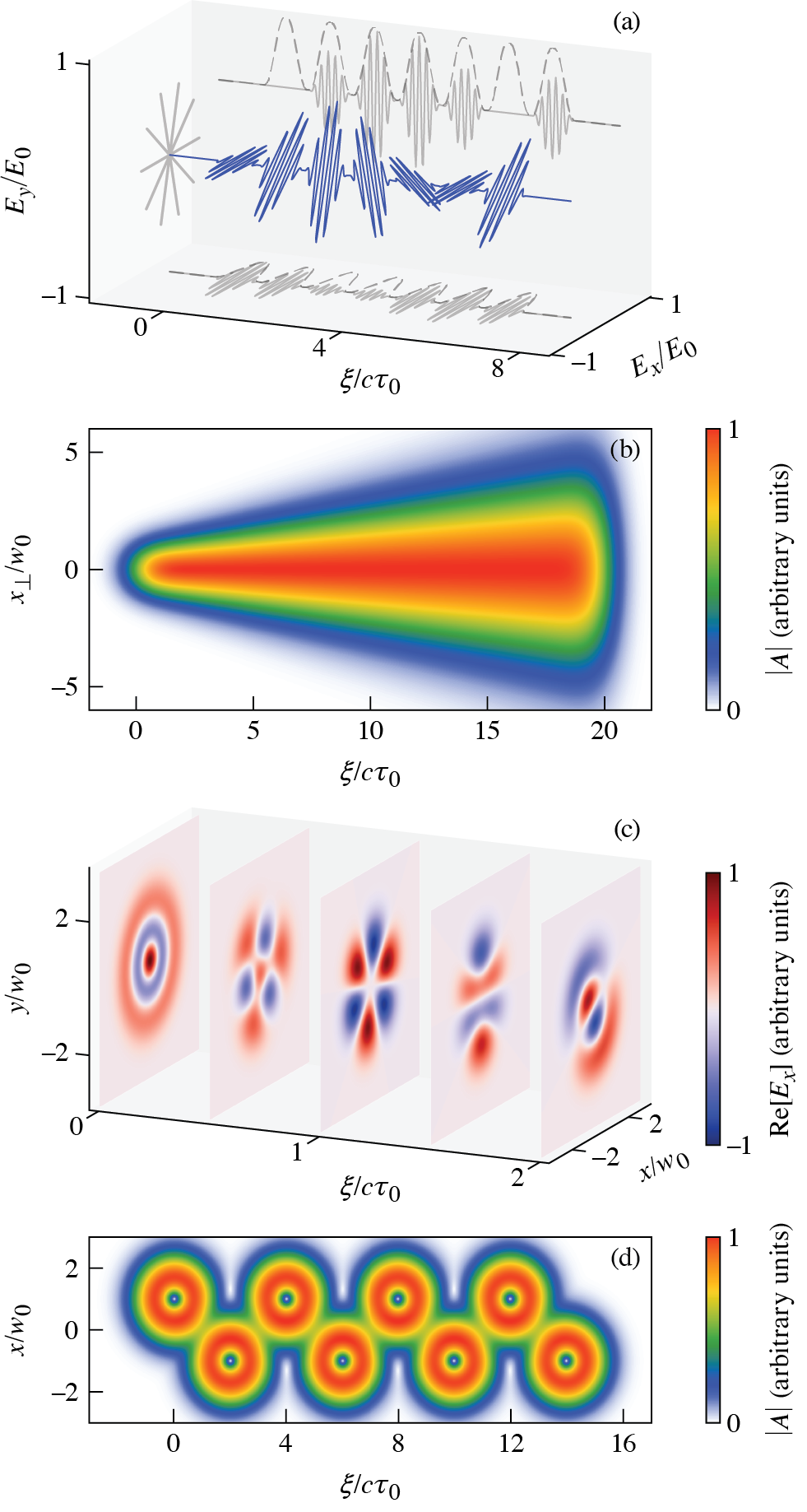}
        \caption{Examples of ASTRL pulses constructed from superpositions of laser pulses focused to the same point ($s_0(\eta) = 0 $): (a) A pulse everywhere linearly polarized, but with a time-dependent polarization angle. The duration of each pulse in the superposition is $\omega_0 \tau_0 = 20$. (b) A laser pulse with a spot size that continuously decreases in time with $k_0 w_{0\text{min}} =40$, $w_{0\text{max}} = 4w_{0\text{min}}$, $\omega_0 \tau_0 = 40 $, and $\Delta_\text{max} = 20 c\tau_0$. (c) Time slices of the transverse electric field of a laser whose angular momentum continuously evolves in time. (d) A lattice of STOVs with  $k_0w_0 = 40$ and $\omega_0 \tau_0 = 20$. All examples are evaluated using Eq. \ref{eq:astrl-general} and shown at the focal point $s=0$.}
        \label{fig:smorgasboard} 
    \end{figure}

    Aside from moving focal points, many applications can benefit from pulses that have a fixed focal point with prescribed spatiotemporal profiles or time-dependent polarization (Fig. \ref{fig:smorgasboard}). Traditional Fourier techniques for temporal pulse shaping, as in Ref \cite{weiner2011ultrafast}, produce pulses described by Eq. \ref{eq:decomp} with a customized $B(\xi)$. Similarly, previous polarization-structuring methods amount to evaluating Eq. \ref{eq:astrl-general} with a prescribed $\hat {\boldsymbol \epsilon}(\eta)$ and $B_{\eta}$, but with a $C_{\eta}$ that is independent of $\eta$. Figure \ref{fig:smorgasboard}(a) shows an example. This particular ASTRL pulse is constructed from a discrete sum of seven pulses, indexed by $\eta \in \{ 0, 1, \hdots 6\}$, that share a common focal point, $s_0(\eta) = 0$. The delay and polarization of each pulse is given by  $\Delta_\eta = -2c\tau_0 \eta$ and $\hat{\boldsymbol{\epsilon}}(\eta) = [ \cos( \pi \eta / 5), \sin(\pi \eta / 5), 0] $ while $C_{\eta}$ is given by Eq. \ref{eq:gaussian-beam}. Evolving polarization structures as in this example have led to advances in nonlinear spectroscopy \cite{silberberg2009quantum}, quantum control \cite{villeneuve2000forced,brixner2004quantum}, nanophotonics \cite{aeschlimann2007adaptive}, and a number of other areas \cite{rubinsztein2016roadmap,misawa2016applications}. More exotic pulses with continuous polarization evolution can also be described using the ASTRL concept. 
    
    A time-dependent spot size can improve the performance of applications including laser wakefield acceleration \cite{bendetti2012quasimatched} and inertial confinement fusion \cite{Igumenshchev2013,froula2013mitigation,huang2016optical}. In laser wakefield acceleration, a spot size that increases in time can stabilize the propagation of an intense laser pulse in a plasma channel \cite{bendetti2012quasimatched}. While in inertial confinement fusion, a spot size that decreases continuously in time can mitigate cross-beam energy transfer \cite{Igumenshchev2013,froula2013mitigation,huang2016optical}. As demonstrated in Fig. \ref{fig:smorgasboard}(b), such a pulse can be described by Eqs. \ref{eq:astrl-general}, \ref{eq:B_delay}, and \ref{eq:gaussian-beam} with a spot size parameterized as $w_0(\eta) = w_{0\text{min}} + \eta ( w_{0\text{max}} - w_{0\text{min}} )$ and the delay $\Delta(\eta) = -\Delta_\text{max} \eta$, where $\eta \in [0,1]$.
    
    The orbital angular momentum and radial mode numbers ($\ell$ and $p$, respectively) of Laguerre-Gaussian modes provide additional degrees of freedom to structure the time-dependence of the transverse profile. Figure \ref{fig:smorgasboard}(c) displays a sequence of time slices in an ASTRL pulse formed by superposing three pulses that partially overlap in time with $(\ell,p) = (0,2), (3,0), (1,1)$. Each pulse is indexed by $\eta \in \{ 0, 1, 2\}$, has a delay $\Delta_\eta = -2c\tau_0 \eta$, and a weight $B_0(\eta) =  1, 0.8, 1$. This construction produces a quasi-continuous transformation in angular momentum. In the first, third, and fifth frames, one pulse dominates, and the transverse profile is nearly a pure Laguerre-Gaussian mode; in the second and fourth frames, two pulses interfere to produce a hybrid mode. Investigations into structured OAM have only recently begun \cite{Zhou:22,rego2019generation,dorney2019controlling}, but its potential for ultrafast probing of chiral systems has already drawn attention \cite{dorney2019controlling}. Many other applications based on OAM \cite{allen1992orbital}, including optical trapping and manipulation \cite{he1995optical,gahagan1996optical,prentice2004manipulation,lorenz2007vortex}, imaging \cite{furhapter2005spiral,furhapter2005spiral2,maurer2011spatial}, quantum optics \cite{mair2001entanglement,malik2016multi}, nonlinear optics \cite{dholakia1996second}, and laser-plasma interactions \cite{mendoncca2009stimulated,ali2010inverse,gariepy2014creating,vieira2014nonlinear,vieira2016high,vieira2016amplification,shi2018magnetic,nuter2022raman}, may also benefit from added control over its structure.  

    As a final example, the ASTRL concept can be used to construct STOVs. In fact, the vacuum STOV solution introduced in Ref. \cite{jhajj2016spatiotemporal}, 
    \begin{equation}
        \begin{split}
            A( \mathbf x_\perp,s, \xi  ) &= \left[ \frac{\xi}{c\tau_0}   \pm \frac{ix}{w(s)} \left( \frac{-q^*(s)}{q(s)} \right)^{1/2} \right] 
            \\ & \frac{z_R}{q(s)} \exp\left[ - \frac{ik_0  x_\perp^2}{2 q(s)} \right] \exp\left[ -\left( \frac{\xi}{c\tau_0}\right)^2 \right],      
        \end{split}
        \label{eq:stov}
    \end{equation}
    is a superposition of two separable solutions as in Eq. \ref{eq:decomp}, where the $C_\eta$ functions are Hermite-Gaussian modes of order $(0,0)$ and $(1,0)$. Here the $s$-dependence has been included, $w(s) \equiv w_0 \sqrt{ 1 + (s/z_R)^2}$, and $q^*(s)$ is the complex conjugate of $q(s)$. The generalization of Eq. \ref{eq:stov} to higher-order vortex topologies (see Ref. \cite{hancock2021stov}) can also be expressed as a superposition of separable solutions. Figure \ref{fig:smorgasboard}(d) illustrates a novel STOV ``lattice" assembled by superposing solutions in the form of Eq. \ref{eq:stov}. The solutions are indexed by $\eta \in \{ 0, 1, \hdots 7\}$ with delays $\Delta_\eta = -2 c\tau_0 \eta$, transverse displacements alternating between $x_0 = \pm w_0$, and polarities (the sign in Eq. \ref{eq:stov}) alternating between $\pm 1$.
     
    The ASTRL description is particularly useful for simulations of structured light and its applications. Previous simulations of flying focus pulses required a Fresnel integral to propagate the field from the optic plane of a specific experimental design to the simulation plane. In contrast, the ASTRL description enables direct field initialization in the simulation plane, which simplifies implementation. Further, by decoupling the structured light-matter interaction physics from the details of a particular optical configuration, the ASTRL description expedites prototyping of new concepts. For example, the axiparabola-echelon pair used in Ref. \cite{Palastro2020} would not need to be redesigned for every simulation. An optical configuration capable of delivering the desired pulse can be designed or implemented in a simulation after a concept shows promise (see Supplemental).  
    
    As demonstrated by the examples above, an ASTRL pulse may generally exhibit spatiotemporal coupling despite being assembled from pulses with no spatiotemporal coupling (i.e., Eq. \ref{eq:decomp}). The separability of Eq. \ref{eq:decomp} was derived by reducing the full wave equation (Eq. \ref{eq:envelope-galilean}) to the paraxial wave equation (Eq. \ref{eq:envelope-galilean-approx}). The validity of the approximations leading to Eq. \ref{eq:envelope-galilean-approx} may be determined by evaluating derivatives of the approximate solution $ A(\mathbf x_\perp,s,\xi) = z_R / q(s) \exp[ - ik_0 x_\perp^2 / 2q(s) ] \exp[ - (\xi/c\tau_0)^2  ]$. At any location, $|\partial_s^2 A | \ll | 2 i k_0 \partial_s A|$ when $ k_0 w_0 \gg 1 $; this is the usual paraxial approximation, which fails only for tightly focused laser pulses. On the other hand, $|2 \partial_\xi \partial_s A | \ll | 2 i k_0 \partial_s A |$ when $\omega_0 \tau_0 \gg 1$. This condition only fails for pulses with durations less than a few-cycles. These two conditions, i.e., $k_0 w_0 \gg 1$ and $\omega_0 \tau_0 \gg 1$, place constraints on $C_{\eta}$ and $B_{\eta}$, respectively.  
    
    For an envelope that is initially separable at some $s=s_i$, spatiotemporal coupling will develop beyond a range $|s-s_i| \gtrsim L_\text{stc} \equiv \frac 1 2 c \tau_0 k_0^2 w_0^2$ because of the $\partial_\xi \partial_z $ term in Eq. \ref{eq:envelope-galilean} (see Supplemental). There are two common scenarios that can be considered. In the first scenario, every pulse comprising an ASTRL pulse is separable at its focus. Because each pulse only contributes significantly within a Rayleigh range of its focus, the condition on the accuracy of Eq. \ref{eq:decomp} can be expressed as $ L_\text{stc} \gtrsim z_R $ or, equivalently, $\omega_0 \tau_0 \gtrsim 1$. In the second scenario, which is more applicable to field initialization in a simulation, every pulse is separable at a specified distance $L_\eta$ from focus. The condition on the accuracy of Eq. \ref{eq:decomp} is then given by $ L_\text{stc} \gtrsim \max_\eta L_\eta $. For example, in Fig. 1(a-b) $\max_\eta L_\eta = \frac 1 2 L_\text{stc} $, which contributes to the slight discrepancy at the end of the focal region. However, even if $ L_\text{stc} \lesssim \max_\eta L_\eta $ and spatiotemporal coupling develops in the constituent pulses, the qualitative features of the desired electromagnetic structure would remain intact. 
    
    The generality of the ASTRL concept extends well beyond Eq. \ref{eq:astrl-general} and the examples presented here. The time-dependence of any combination of parameters, such as the longitudinal focal point and OAM, can be simultaneously structured. Further, the ASTRL concept can be extended to accommodate multi-color pulses \cite{kim2007thz,edwards2014enhanced}, superpositions of pulses with vector polarization \cite{zhan2009cylindrical,wang2010new,Milione2011,naidoo2016controlled}, and pulses structured in the spatiospectral domain. In fact, using a wavelet transform one can show that arbitrary monochromatic laser fields may be decomposed as a generalized version of Eq. \ref{eq:astrl-general} (see Supplemental). 
    
    The ASTRL formalism provides a framework for constructing laser pulses with desired spatiotemporal structure. An ASTRL pulse is synthesized using superpositions of known solutions to Maxwell's equations, such as an ensemble of traditional laser pulses. The flexibility of the ASTRL concept was illustrated with several examples, including pulses with moving focal points, dynamic polarization, evolving angular momentum, and nontrivial topological structure. In addition to facilitating the creation of new and previously unimagined electromagnetic structures, the ASTRL formalism simplifies the injection of structured light in simulations. New pulses based on this concept may enable or enhance techniques in a range of scientific disciplines, including microscopy, non-linear optics, quantum optics, and laser-plasma interactions.

\begin{acknowledgments}
    The authors graciously acknowledge 
	C. Joshi,
	D. H. Froula,
	M. Vranic,
	E. P. Alves,
    SJ Spencer, 
	A. Di Piazza,
	M. Formanek,
	T. Carbin,
	and
	C. Barkan for their lucid insights. 

    Work supported by the National Science Foundation award 2108970, the Department of Energy contracts DE-SC00215057 and DE-SC0010064, the Scientific Discovery through Advanced Computing (SciDAC) program through a Fermi National Accelerator Laboratory (FNAL) subcontract No. 644405, and the Unversity of Rochester, Laboratory for Laser Energetics. Computer simulations were performed on NERSC’s Cori cluster (account m1157). J. Pierce was supported by a scholarship from the Directed Energy Professional Society.

\end{acknowledgments}

\bibliography{bib}

\end{document}


\preprint{OSESPA ARXIV}

\title{Supplemental: Arbitrarily Structured Laser Pulses\\}

\author{Jacob R. Pierce}
    \email{jacobpierce@physics.ucla.edu}
    \affiliation{Department of Physics and Astronomy, University of California, Los Angeles, California 90095, USA}
 
 \author{John P. Palastro}%
    \email{jpal@lle.rochester.edu}
    \affiliation{ University of Rochester, Laboratory for Laser Energetics, Rochester, New York 14623, USA}


\author{Fei Li}
    \affiliation{Department of Physics and Astronomy, University of California, Los Angeles, California 90095, USA}

\author{Bernardo Malaca}
    \affiliation{GoLP/Instituto de Plasmas e Fusão Nuclear, Instituto Superior Técnico, Universidade de Lisboa, Lisbon 1049-001, Portugal}

\author{Dillon Ramsey} 
    \affiliation{ University of Rochester, Laboratory for Laser Energetics, Rochester, New York 14623, USA}


\author{Jorge Vieira}
    \affiliation{GoLP/Instituto de Plasmas e Fusão Nuclear, Instituto Superior Técnico, Universidade de Lisboa, Lisbon 1049-001, Portugal}

    
\author{Kathleen Weichman} 
    \affiliation{ University of Rochester, Laboratory for Laser Energetics, Rochester, New York 14623, USA}
    
\author{Warren B. Mori}
    \affiliation{Department of Physics and Astronomy, University of California, Los Angeles, California 90095, USA}


\date{\today}

\newcommand{\Bavg}{\ensuremath{\langle \beta_z \rangle} }                    

\newcommand{\gamPsqAvg}{\ensuremath{\langle \gamma_\perp^2 \rangle} }        

\newcommand{\PZavg}{\ensuremath{\langle u_z \rangle} }                       

\newcommand{\gamAvg}{\ensuremath{\langle \gamma \rangle} }                   

\newcommand{\appropto}{\mathrel{\vcenter{                                    
  \offinterlineskip\halign{\hfil$##$\cr
    \propto\cr\noalign{\kern2pt}\sim\cr\noalign{\kern-2pt}}}}}


\maketitle


\section{A Conceptual Approach for Synthesizing ASTRL Pulses in Practice}

    \label{sec:discrete-astrl} 
    
    Wavelength-multiplexed communication and power scaling have motivated decades of progress in methods for combining continuous-wave laser beams \cite{Tomlinson:77,leger1993external,fan2005,brignon2013coherent}. Conceptually, $N$ independent beams or $N$ copies of one beam are individually amplified and then recombined into a single beam with $N$ times the original power. Splitting and recombination are achieved using standard optics, such as diffraction gratings, fibers, and beamsplitters. More recently, these methods have been applied to ultrashort laser pulses as a path towards pulses beyond the petawatt class \cite{Zhou:07,seise2010coherent,pupeza2010power,seise2011coherently,daniault2011coherent,Klenke:11,Kienel:13,klenke2013530,Klenke:14,zhou2015coherent,Zhou:17,Klenke:18,Fsaifes:20,wang2021stabilization} and generation of high-harmonic attosecond pulses \cite{dorney2019controlling}.

    Divided pulse amplification techniques can be extended to produce ASTRL pulses. In addition to being amplified, the separated pulses can be independently focused, delayed, polarization-rotated, or altered in other ways before being recombined. While alignment and timing may be challenging, divided pulse manipulation would provide unprecedented control over the structure of the combined pulse and enable a broad class of pulses described by Eq. 7. 
    
    This approach to synthesizing ASTRL pulses is based on approximating the integral in Eq. 7 as a discrete sum of pulses. To illustrate this, consider the $\beta_f = 1.05$ flying focus example shown in Fig. 1. Figure \ref{fig:discrete-comparison} demonstrates that $20$ pulses are more than sufficient to approximate the integral when $L=10z_R$. 

    \begin{figure}[h!]
    \includegraphics{images/Figure3.pnga}
        \caption{(a,b) Discrete approximations of Eq. 7 for the $\beta_f = 1.05$ flying focus pulse shown in Fig. 1 with $N=10$ and $N=20$ pulses, respectively; the continuum is well approximated by 20 pulses, or 2 pulses per $z_R$. (c) Peak on-axis amplitude for the $N=10$ and $20$ cases, showing oscillation for $N=10$. (d) Relative amplitude oscillation, defined as $ (\max{|A|} - \min{|A|})/(\max{|A|} + \min{|A|})$ over the interval $[-z_R, z_R]$. The relative oscillation decreases exponentially with $N$, which determines the number of pulses required for a given tolerance.}
    \label{fig:discrete-comparison}
    \end{figure}
    
    More generally, the minimum number of laser pulses required to approximate Eq. 7 can be estimated by ensuring that the sum-total duration of all constituent pulses is greater than the duration of the entire ASTRL pulse $T$, i.e., $ N \gtrsim T / \tau_0$. For a flying focus pulse, the effective duration of the intensity peak, $\tau_{\text{eff}} = (1-\beta_f)z_R/ \mathrm{v}_f$ \cite{Simpson:22}, satisfies $\tau_{\text{eff}}/T = z_R/L$, which implies that $N \gtrsim (1-\beta_f)L / \mathrm{v}_f \tau_0$ pulses are required. Note, however, that for certain applications, discrete separation of the constituent pulses as in the $N=10$ case of Fig. \ref{fig:discrete-comparison} may also be desirable. 
    
\section{Length Scale for Emergence of Spatiotemporal Coupling}
    
    Over a long enough propagation path, the $2 \partial_s \partial_\xi$ term omitted in Eq. 5 will lead to the development of spatiotemporal coupling in an initially separable envelope. The length scale over which this coupling develops can be estimated using geometric optics. Consider a pulse initialized by Eq. 6 at $s = -s_0$ with a focal point at $s=0$. The distance from a radial location $r$ to the focal point is given by $d(r) = \sqrt{ r^2 + s_0^2 }$. The separable form of Eq. 6 begins to break down when the delay between the center and outer radii of the pulse becomes comparable to its duration $\tau_0$. For an initial Gaussian transverse profile, the spot size at $s = -s_0$ is $w(-s_0) \approx w_0 s_0 / z_R = 2 s_0 / k_0 w_0 $, where $w_0$ is the spot size at focus. Setting $r = w$ provides the delay $ \Delta \sim \sqrt{ (2s_0 /k_0 w_0 )^2 + s_0^2 } - s_0 \sim 2 s_0 / (k_0 w_0)^2 $. Equating $\Delta \gtrsim c\tau_0$ demonstrates that Eq. 6 breaks down for propagation distances 
        \begin{equation}
            s_0 \gtrsim L_\text{stc} \equiv \frac 1 2 c\tau_0 k_0^2 w_0^2. 
            \label{eq:s_stc}
            \tag{S1}
        \end{equation}
    For most pulses of interest, $\omega_0 \tau_0 \gg 1$ so that $L_\text{stc} \gg z_R$, i.e., Eq. 6 remains valid over distances much longer than a Rayleigh range.
    
    Equation \ref{eq:s_stc} can also be derived using a multi-scale expansion of the operators in Eq. 4, where it is explicitly assumed that the mixed derivative term is a first-order correction. The $\partial_s $ operator on the full solution $A$ may be expanded as $ \partial_s \approx \partial_{s0} + \partial_{s1}$, where $ \partial_{s0} \gg \partial_{s1} $. To lowest order, $k_0\partial_{s0}$ balances $\nabla_{\perp}^2$ and one finds the scaling, $\partial_{s0} \sim z_R^{-1}$. At first order, $k_0\partial_{s1}$ balances $\partial_{s0} \partial_\xi $, providing the scaling relation $  k_0 / L_\text{stc} \approx 1/c\tau_0 z_R $. Upon solving for $L_\text{stc}$, this simplifies to Eq. \ref{eq:s_stc}.

\section{Separable Envelope Decomposition of Arbitrary Laser Fields}
        
        Consider an electric field with characteristics of a monochromatic laser pulse, i.e., a power spectrum peaked at a single $\omega_0$ such that $\omega_0\tau_0 \gtrsim 1$, but otherwise arbitrary structure. It is possible to show that such an arbitrary field may be decomposed as a superposition of separable pulses through a generalization of Eq. 7. In particular, the envelope may be exactly decomposed as
        \begin{equation}
            A(\mathbf x_\perp,  s, \xi ) = C_\psi^{-1} \int_\mathbb R \int_\mathbb R \frac{da \, db}{a^2 |a|^{1/2}}  C^{(a,b)}(\mathbf x_\perp,s) \psi \left( \frac{\xi - b }{a} \right) 
            \tag{S2}
            \label{eq:wavelet-decomp}
        \end{equation}
        where $ C_\psi^{-1}$ is a constant, $\psi$ is a wavelet mother function, and
        \begin{equation}
            C^{(a,b)}(\mathbf x_\perp,s) = \frac 1 {|a|^{1/2}} \int_\mathbb R d\xi \, A( \mathbf x_\perp, s, \xi ) \psi^* \left( \frac{\xi - b }{a} \right) 
            \tag{S3}
        \end{equation}
        is the wavelet transform of $A(\mathbf x_\perp, s,\xi)$ with respect to $\xi$ for fixed $\mathbf x_\perp$ and $s$ \cite{Debnath2015}. For example, one may use the Ricker wavelet $\psi(g) = \psi_0 \exp[ -g^2 / 2] ( 1 - g^2) $. In Eq. \ref{eq:wavelet-decomp}, $a$ and $b$ are respectively the length and delay of each constituent pulse. The properties of each $C^{(a,b)}(\mathbf x_\perp,s)$, such as spot size and focal point, may depend on $a$ and $b$.
        
        Equation \ref{eq:wavelet-decomp} is a superposition of separable pulses at each slice in $s$. Given initial data for the envelope $A(\mathbf x_\perp, 0, \xi)$ at some slice $s=0$, the decomposition may be used to approximately evolve the full solution $A(\mathbf x_\perp, s, \xi)$ using Eq. 6 as        
        \begin{equation}
            A( \mathbf x_\perp, s, \xi ) \approx C_\psi^{-1} \int_\mathbb R \int_\mathbb R \frac{da \, db}{a^2 |a|^{1/2}} C_\text{parax}^{(a,b)}( \mathbf x_\perp,s ) \psi \left( \frac{\xi - b }{a} \right)  
            \tag{S4}
            \label{eq:astrl-most-general}
        \end{equation}
        where $C_\text{parax}^{(a,b)}( \mathbf x_\perp,s )$ are solutions to the paraxial wave equation 
        $\left[ 2ik_0 \partial_s + \nabla_\perp^2 \right]  C_\text{parax}(\mathbf x_\perp,s) = 0 $ 
        with initial condition $ C_\text{parax}(\mathbf x_\perp,0) = C^{(a,b)}(\mathbf x_\perp,0)$. The properties of $C_\text{parax}^{(a,b)}$  determine the range of $s$ over which Eq. \ref{eq:astrl-most-general} remains valid. For example, if $C_0^{(a,b)}(\mathbf x_\perp)$ resembles a Gaussian beam with spot size $w_0(a,b)$ for each $(a,b)$, then Equation \ref{eq:astrl-most-general} remains valid when 
        \begin{equation}
            |s| \lesssim \min_{a,b} L_\text{stc} (a,b) = \min_{a,b}  a k_0^2 w_0(a,b)^2,            \tag{S5}
        \end{equation}
         where the minimum is taken over $(a,b)$ with significant contribution to the integral in Eq. \ref{eq:wavelet-decomp}. Note that the specific choice of wavelet is not important, but the wavelet determines the conditions required for validity of Eq. \ref{eq:astrl-most-general}. 
        
        Equation \ref{eq:astrl-most-general} generalizes Eq. 7 and shows that any fully coherent laser pulse solution may be approximately decomposed as a superposition of pulses with separable envelopes. In this sense, all monochromatic laser pulses can be constructed as ASTRL pulses. However, we restrict the scope of ASTRL pulses to include pulses that have been intentionally designed to exhibit a particular structure either through Eq. 2 or Eq. 7.

\bibliography{bib}